# MicroRNA Systems Biology

# Edwin Wang

1. Biotechnology Research Institute, National Research Council of Canada, 6100 Royalmount Avenue, Montreal, Quebec, Canada, and 2. Center for Bioinformatics, McGill University, Montreal, Quebec, Canada

Email: Edwin.Wang@cnrc-nrc.gc.ca
Fax : 1-514-496-5143
Tel : 1-514-496-0914

More similar work can be found at: www.bri.nrc.ca/wang

**Abstract**

Recently, microRNAs (miRNAs) have emerged as central posttranscriptional regulators of gene expression. miRNAs regulate many key biological processes, including cell growth, death, development and differentiation. This discovery is challenging the central dogma of molecular biology. Genes are working together by forming cellular networks. It has become an emerging concept that miRNAs could intertwine with cellular networks to exert their function. Thus, it is essential to understand how miRNAs take part in cellular processes at a systems-level. In this review, I will first introduce basic knowledge of miRNAs and their relations to heart disaeses and cancer, highlight recently dicovered functions such as filtering out gene expression noise by miRNAs. I will aslo introduce basic concepts of cellular networks and interpret their biological meaning in such a way that the network concepts are digested in a biological context and are understandable for biologists. Finally, I will summarize the most recent progress in understanding of miRNA biology at a systems-level: the principles of miRNA regulation of the major cellular networks including signaling, metabolic, protein interaction and gene regulatory networks. A common miRNA regulatory principle is emerging: miRNAs preferentially regulated the genes that have high regulation complexity. In addition, miRNAs preferentially regulate positive regulatory motifs, highly connected scaffolds and the most network downstream components of cellular signaling networks, while miRNAs selectively regulate the genes which have specific network structural features on metabolic networks.

## 1.1 Introduction

According to the central dogma of molecular biology, RNAs are passive messengers and only take charge of transferring genetic information, or carrying out DNA instructions, or code, for protein production in cells. However, this central dogma is getting challenged by the recent findings that tiny fragments of noncoding RNA, typically ~22 nucleotides in length, namely microRNA (miRNA), are able to negatively regulate protein-coding genes by interfering with mRNA's original instructions. Recent studies indicate that miRNAs have emerged as central posttranscriptional repressors of gene expression. miRNAs suppress gene expression via imperfect base pairing to the 3′ untranslated region (3′UTR) of target mRNAs, leading to repression of protein production or mRNA degradation (Bartel, 2004; Carthew, 2006; Valencia-Sanchez et al. 2006). These noncoding regulatory RNA molecules have been found in diverse plants, animals, some viruses and even algae species and it now seems likely that all multicellular eukaryotes, and perhaps some unicellular eukaryotes, utilize these RNAs to regulate gene expression.

Some researchers claimed that the human genome might encode more than 1,000 miRNAs (Bentwich et al. 2005), however, a recent sequencing survey of miRNA expression cross 26 distinct organ systems and cell types of human and rodents validated that only over 300 miRNAs are present in humans and/or rodents (Landgraf et al. 2007). Computational predictions indicate that thousands of genes could be targeted by miRNAs in mammals (John et al. 2004; Krek et al. 2005; Lewis et al. 2003; Rajewsky, 2006). Experimental analysis revealed that 100 to 200 target mRNAs are repressed and destabilized by a single miRNA (Krutzfeldt et al. 2005; Lim et al. 2005; Yu et al. 2007a). It is estimated that more than one third of human genes are potentially regulated by miRNAs. These findings suggest that miRNAs play an integral role in genome-wide regulation of gene expression.

miRNAs regulate many key biological processes, including cell growth, death, development and differentiation, by determining how and when genes turn on and off. Animals that fail to produce certain mature miRNAs are unable to survive or reproduce (Bernstein et al. 2003; Forstemann et al. 2005; Ketting et al. 2001; Wienholds et al. 2003; Cao et al. 2006; Plasterk, 2006; Shivdasani, 2006). Thus, a single, malfunctioning microRNA can be sufficient to cause cancer in mice (Costinean et al. 2006). These discoveries offer new insight into another layer of gene regulation and at the same time underscore the powerful role that these tiny snippets of non-coding RNA play in cells. These discoveries indicate that it is no longer the genes, or mRNAs themselves that held the most intrigue, but the miRNAs that influence their behavior and the result that such gene regulation process produces. Thus, miRNA has become an important force in biology.

In cells, genes are not isolated and do not independently perform a single task; instead, genes are grouped to collaborate and carry out some specific biological function. This collaborative effort of genes indicates that genes are working together in a cell, and we can formulate this conceptually depicting the various interactions as cellular networks. miRNAs can be regarded as regulators that regulate all kinds of cellular networks and can also be treated as network components that are involved in many cellular functions. Thus, it is imperative to understand how miRNAs take part in cellular processes at a systems-level. In this review, I will first introduce basic knowledge of miRNAs, their biogenesis, functions, and relations to heart disaese and cancer, and then introduce the basic concepts and biological meaning of cellular networks. Network concepts are borrowed from mathematics and graph theory, however, I will digest the concepts in a biological context and make them understandable for biologists. Finally, I will summarize the most recent progress in understanding of miRNA biology at a systems-level: the principles of miRNA regulation of the major cellular networks, including signaling, metabolic, protein interaction and gene regulatory networks.

**1.1.1 miRNA biogenesis**
miRNAs as posttranscriptional regulatory molecules were first discovered to regulate expression of partially complementary mRNAs in *Caenorhabditis elegans* (Lee et al. 1993; Wightman et al. 1993; Moss et al. 1997). miRNAs are encoded in either intergenic regions of genomes or within introns of known protein-coding genes. miRNAs are transcribed by RNA polymerase II as long precursor transcripts, which are called primary miRNAs (pri-miRNAs). The pri-miRNAs are capped and polyadenylated, and can reach several kilobases in length (Cullen, 2005; Kim, 2005). A single pri-miRNA might contain one, or up to several miRNAs. Several sequential steps of transcript processing are required to produce mature miRNAs from pri-miRNAs. In the nucleus, the microprocessor complex in which the major components are the RNase-III enzyme Drosha and its partner DGCR8/Pasha (Denli et al. 2004; Gregory et al. 2004; Landthaler et al. 2004), which are initially recognize pri-miRNAs and then excise the stem–loop hairpin structure that contains the miRNA, a 60–80 nucleotide intermediate knowen as precursor miRNA (pre-miRNA) with pri-miRNAs. Exportin-5, a nuclear export factor, recognizes and transports the pre-miRNAs to cytoplasm (Yi et al. 2003; Bohnsack et al. 2004; Lund et al. 2004). In the cytoplasm, Dicer, a second RNase-III enzyme, cleavages the pre-miRNAs to generate double-stranded 18–24 nucleotide-long RNA molecules – miRNAs (Bernstein et al. 2001; Grishok et al. 2001; Hutvagner et al. 2001; Ketting et al. 2001; Knight and Bass, 2001). RNA-induced silencing complex (RISC), the core component of which is the Argonaute protein (Kim, 2005), incorporates one of these two strands – the guide strand of miRNAs. Finally, the miRNA guides the RISC complex to the target mRNA to suppress gene expression via imperfect base pairing to the 3′UTR of target mRNAs, leading to repression of protein production and, in some cases, mRNA degradation (Bartel, 2004; Carthew, 2006; Valencia-Sanchez et al. 2006).

**1.1.2 Biological functions of miRNAs**

**1.1.2.1 miRNA emerges as a central regulator for development**
Several reports indicated that miRNAs repress a large set of targets so that the targets are expressed at low levels in the miRNA-expressing cells (Krutzfeldt et al. 2005; Lim et al. 2005; Yu et al. 2007a). This might offer a second layer of regulation to reinforce transcriptional controls at posttranscriptional level. A number of lines of evidence suggested that miRNAs are involved in regulating developmental processes. We conducted a genome-wide survey of transcription factor binding sites in the promoter regions of human genes and found that developmental genes are significantly regulated by more transcription factors (Cui et al. 2007a). Furthermore, we showed that the more transcription factors a gene is regulated by, the more miRNAs that gene is regulated by (Cui et al. 2007a). Certain miRNAs have been suggested to be essential regulators for developmental programs (Giraldez et al. 2005). For instance, without miR-430, zebrafish embryos develop defects, which can be rescued and complemented by supplying miR-430 (Giraldez et al. 2005). Genes in this process by miR-430 seem to be direct miR-430 targets based on miRNA seed matches, are misregulated in the absence of miR-430 (Giraldez et al. 2006). Another example comes from the study of *C. elegans* miRNAs, lin-4 and let-7. Without lin-4, *C. elegans* is unable to make the transition from the first to the second larval stage due to a differentiation defect, which is caused by a failure to posttranscriptionally repress the lin-14 gene, which is the target gene of lin-4 (Lee et al. 1993; Wightman et al. 1993). Similarly, without let-7, a failure of larval-to-adult transition was observed (Reinhart et al. 2000). It is known that lin-41, hbl-1, daf-12 and the forkhead transcription factor pha-4 are the direct targets of let-7 during this transition (Abrahante et al. 2003; Grosshans et al. 2005; Slack et al. 2000).

**1.1.2.2 miRNAs are involved in cell proliferation and apoptosis**
miRNAs have been shown to regulate key genes for tumorigenesis and cancer progression, which coordinately controls cell proliferation and apoptosis. For instance, miRNA let-7 promotes tumorigenesis by regulation KRAS and NRAS transcripts (Johnson et al. 2005). miRNAs are known to regulate pathways controlled by genes like p53, MYC and RAS. Furthermore, miR17-92 cluster has been shown to be able to act as a functional switch between cell proliferation and apoptosis.

**1.1.2.3 miRNAs act as regulators for noise filtering and buffering**
Eukaryotic cells are noisy environments in which transcription often occurs in a bursting manner, causing the number of mRNAs per cell to fluctuate significantly (Blake et al. 2006; Golding et al. 2005; Raj et al. 2006). Moreover, such fluctuations can propagate through the network, e.g., fluctuations in the level of an upstream transcription factor can significantly induce the expression fluctuations of downstream genes (Pedraza and van, 2005; Rosenfeld et al. 2005). In positive regulatory loops, noise or stochastic fluctuations of gene transcripts and protein molecules leads to randomly switching cell phenotypes in yeast, while a negative regulator adding in the positive regulatory loops often helps reducing such noise in biological systems and making a robust decision for cell development (Acar et al. 2005). Because miRNAs can tune target protein levels more rapidly at the posttranscriptional level, they might significantly shorten the response delay and, in turn, provide more effective noise buffering. The miRNA, miR-17 might play a role in preventing noise-driven transition from apoptosis to cell proliferation. c-Myc and E2F1 are known to reciprocally activate transcription of one another, establishing a positive feedback circuit (Fernandez et al. 2003; Leone et al. 1997). This architectural structure of the circuit makes it possible for miRNAs to support a shift from apoptosis toward proliferation by repressing E2F1. Expression of E2F1 promotes G1 to S phase progression by activating genes involved in cell cycle (Bracken et al. 2004). High expression of E2F1, however, is sufficient to induce apoptosis (Matsumura et al. 2003; Johnson et al. 1994a; Johnson et al. 1994b). In the absence of additional regulatory mechanisms, this circuit might be expected to overactivate E2F1, leading to apoptosis. When c-Myc simultaneously activates E2F1 transcription and miR17-92 cluster, which in turn negatively regulates E2F1. This might promote a proliferative signal but not a apoptotic signal. Another example is the fly miR-9a, which is suggested to set up a 'threshold' for signals in a positive feedback loop, so that it can filter out noise (Li et al. 2006). Without miR-9a, flies produce extra sense organs (Li et al. 2006). During fly sensory organ development, a fly gene, *senseless* expression is activated by proneural proteins and feedbacks positively to reinforce proneural gene expression. If *senseless,* the target of miR-9a, is highly expressed, the defects mentioned above occur. miR-9a has been suggested to set a threshold that *senseless* expression has to overcome to induce the normal developmental program. In agreement with these findings, we found that cross-species expression divergences of miRNA target genes are significantly smaller than those of other genes (Cui et al. 2007b). Similar observations have been found between human and chimpanzee, between human and mouse, *Drosophila* species, and *D. melanogaster* and *D. simulans* (Cui et al. 2007b). These results suggest that miRNAs might provide a genetic buffer to constrain gene expression divergence. We showed that miRNAs preferentially regulate positive regulatory loops (Cui et al. 2006). It is possible that miRNAs serve to buffer stochastically fluctuating expression of genes in positive regulatory loop, in turn, provide a common mechanism in buffering gene expression noise. Buffering by miRNAs decreases the detrimental effects of errors in gene regulation. miRNA buffering might also provide a way for silence-accumulating mutations without being subjected to selective forces and thus might contribute to evolvability.

**1.1.2.4 miRNAs might contribute to maintaining tissue identity**
We conducted a genome-wide analysis of the expression profiles of mRNA targets in human, mouse and *Drosophila* (Yu et al. 2007a). We found that the expression levels of miRNA targets are significantly lower in all mouse mature tissues and *Drosophila* later life stages than in the embryos. These results indicate that miRNAs might play roles in determining the timing of tissue differentiation during larva period of *Drosophila* development and maintaining the tissue identity during the adulthood.

**1.1.3 miRNAs in human disease**
**1.1.3.1 miRNAs and heart diseases**
Loss- or gain-of-function of specific miRNAs appears to be a key event in the genesis of many diverse diseases. Recently, an interesting question how miRNAs influence heart development and disease has been addressed. Four recent papers highlighted the role of miRNAs in the heart. These reports showed that miRNAs are essential for heart development and regulating the expression of genes which take part in cardiac function in vivo: the conductance of electrical signals, heart muscle contraction, and heart growth and morphogenesis.

Yang et al. reported a function for miR-1 in heart conductivity. miR-1 levels were positively correlated with coronary artery disease and rats after cardiac infarction (Yang et al. 2007). Loss-of-function of miR-1 prevented heart arrhythmia, whereas miR-1 overexpression caused heart arrhythmia in normal and infarcted hearts. They further showed that both gain- and loss-of-function of miR-1 affect conductivity through affecting potassium channels. These results suggest that miR-1 has a prominent effect on the development of cardiac arrhythmia, irregular electrical activity in the heart. In a separate study, Zhao et al. also focused on miR-1 study by creating mice that are deficient in a muscle-specific miRNA, miR-1-2 (Zhao et al. 2007). They showed that the miRNA deficient embryos have cardiac failure and a variety of developmental defects, including pericardial edema and underdevelopment of the ventricular myocardium, increasing in cardiomyocyte proliferation and electrophysiological defects, reducing heart rate and prolonging ventricular depolarization. Interestingly, these phenotypes are similar to the defects during heart development in zebrafish embryos, when miRNAs are non-functional (Giraldez et al. 2005). Both studies identify miR-1 targets that might, at least in part, account for the manifestation of the associated diseases.

In another study, Care et al. found that the muscle-specific miR-133 is a negative regulator of cardiac hypertrophy which is an essential adaptive physiological response to mechanical and hormonal stress and heart size (Care et al. 2007). To understand the molecular mechanism by which miR-133 controls heart size, they showed that Rhoa, Cdc42 and Whsc2 are the direct targets of miR-133. Moreover, van Rooij and colleagues found that the heart-specific miRNA miR-208 also modulates the genes that are controlling the hypertrophic response (van Rooij et al. 2007). The main function of miR-208 seems to be mediating the switch from expression of the heavy chain of α-myosin to that of β-myosin during stress or thyroid-hormone-induced cardiac growth (van Rooij et al. 2007). These results suggest that miR-208 is an important regulator for cardiac growth and gene expression in response to stress and hypothyroidism.

Taken together, it seems clear that miRNAs have an important role in regulating gene expression in the heart. These studies indicate that miRNAs are important during heart development and adult cardiac physiology, and modulate a diverse spectrum of cardiovascular functions in vivo. These findings revealed a level of molecular control of heart physiology that is beyond the well-accepted regulatory role of signaling and transcription factor complexes in the heart. Furthermore, these studies also have implications for understanding complex pathways, e.g., interactions between miRNAs, cell signaling and transcription factors, involved in heart diseases, and lead to potential opportunities in manipulating miRNAs as therapeutic targets.

**1.1.3.2 miRNAs and cancer**
Human cancer studies are always the hotspots in life science research. Much progress in miRNAs and cancer has been made significantly in the past few years. Genome-wide studies of miRNA expression profiling showed that miRNA expression levels are altered in primary human tumors (Calin et al. 2004; Lu et al. 2005). Significant signatures of miRNA expression profiles can be linked to various types of tumors, suggesting that miRNA profiling has diagnostic and perhaps prognostic potential (Lu et al. 2005; Calin and Croce, 2006). Certain miRNAs could be tumor suppressors, because loss of these miRNAs is often associated with cancers. miR-15a and miR-16-1 genes are deleted in most cases of chronic lymphocytic leukemia (Calin et al. 2004). Loss of miRNA let-7 in lung tumors correlates with high RAS protein expression, suggesting that let-7 promote tumorigenesis by regulation popular oncogenes, KRAS and NRAS transcripts (Johnson et al. 2005). miR-372 and miR-373 have been shown to be able to overcome oncogenic

Ras-mediated arrest and, therefore, induced tumorigenesis (Voorhoeve et al. 2006). miR-21 was demonstrated to be consistently upregulated in human glioblastoma tumor tissues, primary tumor cultures and established glioblastoma cell lines relative to normal fetal and adult brain tissue (Chan et al. 2005). Knockdown of miR-21 in glioblastoma cell lines lead to activation of caspases and a corresponding induction of apoptotic cell death. Furthermore, two rececent studies have placed miR-24 family into the p53 tumor spressor network. miR-24, regulating apoptosis and cell proliferation, has become an essential component of the p53 network (He et al. 2007; Raver-Shapira et al. 2007), which is closely associated with cancer.

**1.1.3.3 Single-nucleotide polymorphisms (SNPs) of miRNA binding sites and human diseases**

Chen and Rajewsky used human SNP genotype data (25,000 SNPs) generated in the HapMap and Perlegen projects and mapped onto the 3'-UTR regions of human gene transcripts (Chen and Rajewsky, 2006). They uncovered that SNP density in conserved miRNA sites was lower than in conserved control sites. These results indicate that a large class of computationally predicted conserved miRNA target sites is under significant negative selection. Similarly, we showed the same trend when mining NCBI's dbSNP database (Yu et al. 2007b). These results have implications that SNPs located at miRNA-binding sites are likely to affect the expression of the miRNA target and might contribute to the susceptibility of humans to common diseases. Indeed, naturally occurring polymorphisms in miRNA binding sites have been documented in Tourette's syndrome in humans and muscularity in sheep. (Abelson et al. 2005; Clop et al. 2006)

Motivated by this concept, we explored the effects of miRNA-binding SNPs on cancer susceptibility by genome-wide analyzing the data deposited in NCBI's dbSNP database and human dbEST database (Yu et al. 2007b). Interestingly, we found that the frequencies of the minor alleles (non-target alleles) of the miRNA-binding SNPs are extremely lower. Furthermore, we showed that the average expression level of the non-target alleles of miRNA-binding SNPs is significantly higher than that of the target alleles. Moreover, we identified a set of potential candidates for miRNA-binding SNPs with an aberrant allele frequency present in the human cancer EST database. Finally, we experimentally validated them by sequencing clinical tumor samples.

Although the miRNA inducing disease studies are still in their infancy, miRNAs are known to regulate pathways controlled by genes like p53, MYC and RAS. These findings emphasize the need to integrate the study of miRNA expression and function into other cellular processes such as signaling, gene regulation, and others in order to achieve a complete understanding of this group of disorders. Unraveling miRNA regulatory circuits, even miRNA regulation of cellular networks that are involved in disease development, is challenging, but is essential to gain a comprehensive understanding of the molecular mechanisms of the diseases. Luckily, there have been recent developments in technologies such as microarray and systemic delivery of small RNA systems that allow high-throughput studies the function of miRNAs (Soutschek et al. 2004; Krutzfeldt et al. 2005). These approaches provide promise for understanding miRNA function at systems-level and evenually developing therapeutic strategies based on miRNA overexpression or inhibition.

## 1.2 Basic network concepts and their biological meanings
### 1.2.1 Biological functions and activities are encoded in cellular networks

Traditionally scientists treat cellular events in the view of biological pathways such as signaling pathways and metabolic pathways, study one pathway at a time and then try to compile information from a few pathways together to understand what is going on inside cells. However, enzymes and other proteins, which make up one individual pathway, rarely operate in isolation but "cross-talk" with another pathway's enzymes/proteins to process metabolic flows and signal information. Currently, it is believed that in the cell, no gene is one island. Biological functions are performed by groups of genes which form interdependent interactions and complex cellular networks such as signaling networks, gene regulatory networks and metabolic networks. The biological complexity encoded in cellular networks has become the core of systems biology (Wang et al. 2007). Therefore, a network view, or a systems-level view of cellular events emerges as an important concept.

For a long time, scientists have investigated one gene or one pathway at a time. In 'omics' era, various high-throughput approaches such as genome sequencing technology, microarray technology and proteomic characterization of proteins and complexes have allowed us to gathering vast amounts of data to construct cellular networks. These efforts provide an opportunity to investigate cellular processes at a global level, therefore, it is essential to develop systematic methods for analyzing cellular networks as well as understanding their properties in a biological context. In the past few years, significant progress has been made for the identification and interpretation of the structural properties of cellular networks. This information has shed light on how such properties might reflect the biological meanings and behaviors of cellular networks (Babu et al. 2004; Barabasi and Oltvai, 2004).

### 1.2.2 Types and categories of cellular networks

Four types of cellular networks have been found in cells: protein interaction networks, metabolic networks, gene regulatory networks and signaling networks, which can be further classified into two categories. Protein interaction networks encode the information of proteins and their physical interactions. Protein interaction information in the network ranges from basic cellular machinery such as protein complexes for DNA synthesis, metabolic enzyme complexes, transcription factor complexes, to protein complexes involved in cellular signaling. Simply put, a genome-wide protein interaction network encodes all the protein interaction information cross all biological processes in a cell. A gene regulatory network describes regulatory relationships between transcription factors and the protein-coding genes. Similar to protein interaction networks, a gene regulatory network encodes the gene regulatory information for all biological processes and activities in a cell. Therefore, I have classified protein interaction networks and gene regulatory networks into the first category: general network. The second category of networks identified is: cellular specific network, which encompasses metabolic networks and signaling networks, describing specific cellular activities. A cellular metabolic network collects all the metabolic reactions and metabolic flows, while a signaling network encodes signal information flows and biochemical reactions for signal transductions. Traditionally, both types of information are presented using pathways, e.g., metabolic pathways and signaling pathways. In metabolic network, metabolic pathways are intertwined so that metabolic flows are transferable across different pathways. Certain metabolites can be shared and used by many different pathways, while certain end-product metabolites are able to be produced via bypassing one or several pathways. Signaling networks illustrate inter- and intracellular communications and information processing between signaling proteins. In fact, pathway concept gets fuzzy and many pathways lose their identities in networks (Spirin et al. 2006; Patil and Nielsen, 2005).

### 1.2.3 The structures of cellular networks are 'scale-free'
Cellular networks can be presented as either directed or undirected graphs. Usually in these networks, nodes represent proteins or genes and the links represent the physical interactions between proteins, gene regulatory relationships, or activation/inactivation signaling reaction relationships. Notably, signaling networks contain the most complicated relationships between proteins, e.g., nodes might represent different functional proteins such as kinases, growth factors, ligands, receptors, adaptors, scaffolds, transcription factors and so on, which all have different biochemical functions and are involved in many different types of biochemical reactions that characterize a specific signal transduction machinery.

One common structural property of cellular networks and other real-world networks is their 'scale-free' topology. In a scale-free network, a small number of nodes act as highly connected hubs, whereas most nodes have only a few links. For example, a map describing the air transportation in the United States is a network, in which only a few big airports (hubs) in big cities such as Boston, New York, Chicago and Los Angles have many air routes (links) to other airports, while many small airports just have a few air routes to the nearby big airports. This common structural feature encodes a special property of these networks: they are robust but also very vulnerable to failure and attack (Barabasi and Albert, 1999; Barabasi and Oltvai, 2004). In a scale-free network, random removal of a substantial fraction of the low-linked nodes will make little damage on the network's connectivity, however, targeted removal of the hub nodes will easily disconnect and destroy the network completely, as illustrated by the air transportation map. Disabling big airports (hubs) will wreak havoc in many ways, while damaging a few small airports will have little or no effect on overall air transportation. These features are common in cellular networks too (see next section).

### 1.2.4 Biological insights of hubs in cellular networks
In gene regulatory networks, hub genes are global transcription factors which govern a large number of genes in response to internal and external signals. Indeed, hub transcription factors do control a large spectrum of biological processes through integrative analysis of the yeast gene regulatory network with gene microarray profiles of many different cellular conditions (Luscombe et al. 2004). We showed that hub transcription factors have significantly faster decay rates than non-hub transcription factors in *Escherichia coli* gene regulatory network (Wang and Purisima, 2005). The similar results were observed in yeast recently (Batada et al. 2006). These results suggest that hub transcription factors facilitate a rapid response of the network to external stimuli (Wang and Purisima, 2005). In protein interaction networks, hub proteins take part in many biological processes. Furthermore, hub proteins might be more important for an organism's survival. Removal of hub proteins from an organism would have a much broader effect on the organism than non-hub proteins. Indeed, hub proteins have central positions in cellular networks and are more essential for the organism's survival than other proteins (Babu et al. 2004; Jeong et al. 2001; Wuchty et al. 2006; Wuchty et al. 2003; Calvano et al. 2005). Therefore, the structure of cellular networks not only sheds light on the complex cellular mechanisms and processes, but also gives insight into evolutionary aspects of the proteins involved. Hub proteins are more evolutionarily conserved than non-hub proteins (Saeed and Deane, 2006). One explanation of these phenomena is that hub proteins are subject to selection pressure and constraints, due to their involvements in many biological processes and their multiple interacting protein partners.

In signaling networks, hub proteins are the most usable protein by multiple signaling pathways. They become information exchanging and processing centers of the network. Normally hub proteins are conserved across animals. In signaling networks, certain hub proteins display core genetic buffering properties (Lehner et al. 2006). Similar to signaling networks, hub enzymes are shared by many metabolic pathways. They become the center for metabolic flow exchanging. Generally speaking, in specific cellular networks (singling network and metabolic network), hubs are communication centers for exchanging information, e.g., signal information and metabolic flows, while hubs in general networks (gene regulatory networks and protein interaction networks) are central players involving in broadly biochemical and/or genetic events, e.g., interactions and gene regulations.

### 1.2.5 Network motifs, themes and modules
A group of genes/proteins in a cellular network are able to collaborate to perform certain biological task. We can regard this group of genes as a functional module. Therefore, a complex signaling network can be broken down into distinct regulatory patterns, or network motifs, typically comprised of three to four interacting components capable of signal processing (Babu et al. 2004; Barabasi and Oltvai, 2004). Network motifs are the smallest functional modules in networks. Network motifs are the statistically significant recurring structural patterns or small subgraphs or sub-networks that are found more often in a real network than would be expected by chance (Shen-Orr et al. 2002). In fact, for a long time, these motifs have been known as gene regulatory loops in biology. In gene regulatory networks, three major motifs are found in gene regulatory networks: Single Input Module (SIM), bi-fan and Feedforward Loop (FFL).
It is believed that network motifs have been evolutionarily selected (Conant and Wagner, 2003). In general, positive feedback loops lean to emergent network properties such as ultrasensitivity, bistability and switch-like behavior, while negative feedback loops perform adaptation, desensitization, and preservation of homeostasis (Dekel et al. 2005; Ferrell, 2002; Balazsi et al. 2005; Dekel et al. 2005; Ferrell, Jr., 2002; Luscombe et al. 2004). Another design principle of these motifs is that the transcription factors whose mRNAs have fast decay rates are significantly enriched in these motifs, suggesting that motif structures encode a regulatory behavior: network motifs are able to rapidly respond to internal and external stimuli and decrease cell internal noise (Wang and Purisima, 2005). Both theoretical and experimental studies have shown that network motifs bear distinct regulatory functions and particular kinetic properties that determine the temporal program of gene expression (Mangan et al. 2003). Therefore, the frequencies and types of network motifs with which cells use reveal the regulatory strategies that are selected in different cellular conditions (Balazsi et al. 2005; Kalir, 2001; Wang and Purisima, 2005). For example, FFLs are buffers that respond only to persistent input signals (Mangan and Alon, 2003), which makes them well-suited for responding to endogenous conditions, while the motifs whose key regulator's transcripts have a fast mRNA decay rate are preferentially used for responding to extrogenous conditions (Wang and Purisima, 2005). In signaling networks, network motifs such as switches (Bhalla et al. 2002), gates (Ma'ayan et al. 2005; Blitzer et al. 1998), and positive or negative feedback loops provide specific regulatory capacities in decoding signal strength, processing information and controlling noise (Dublanche et al. 2006).

Network motifs are not isolated in networks, but form large aggregated structures, called network themes that perform specific functions by forming collaborations between a large number of motifs (Zhang et al. 2005). A higher level of aggregation of network themes can be regarded as network modules.

### 1.3 Principles of miRNA regulation of cellular networks

It is currently estimated that miRNAs account for ~ 1% of predicted genes in higher eukaryotic genomes and that up to 10%-30% of genes might be regulated by miRNAs. miRNAs have been shown to have biological functions in many aspects. miRNA targets range from signaling proteins, metabolic enzymes, transcription factors and so on. The diversity and abundance of miRNA targets offer an enormous level of combinatorial possibilities and suggest that miRNAs and their targets appear to form a complex regulatory network intertwined with other cellular networks such as signal transduction networks, metabolic networks, gene regulatory networks and protein interaction networks. It is reasonable to think that miRNAs exert their functions through regulating cellular networks. However, it is unclear how miRNAs orchestrate their regulation of cellular networks and how regulation of these networks might contribute to the biological functions of miRNAs. We have addressed these questions by analyzing the interactions between miRNAs and the major cellular networks in cells. Because many miRNAs are highly conserved, their functions should be advantageous (Pasquinelli et al. 2000; Lagos-Quintana et al. 2001; Lagos-Quintana et al. 2003). Therefore, the principles of miRNA regulation in one organism could be transferable into other closely related organisms.

### 1.3.1 miRNA regulation of cellular signaling networks
### 1.3.1.1 Signaling networks and computational analysis

Specific signaling pathways deploy many different proteins, however, pathways often "talk" each other. This so called "cross-talk" between pathways has been systematically investigated, and an unexpected high numbers of cross-talk events between signaling pathways have been discovered (Lehner et al. 2006; Wang et al. 2007). These studies confirmed that signaling pathways form a complex network to process information (Wang et al. 2007). Cellular signaling networks encode inter- and intracellular communications and information processes between signaling proteins. The components of cellular signaling networks, mainly composed of proteins, are activated or inhibited in response to specific input stimuli and, in turn, serve as stimuli for further downstream proteins. Cellular signaling network is the primary complex cellular system to responding stimuli, signals and messages from other cells and environment. Once a cell receives signals, it processes the information, e.g., signal amplification and noise filtration, and finally the signals reach to transcription factors so that the signaling network triggers the responses of gene regulatory networks. Therefore, a signaling network is the most important complex system in processing the early extra- and intra-cellular signals in a cell. Cells use signaling networks, as a sophisticated communication system, to perform a series of tasks such as growth and maintenance, cell survival, apoptosis and development.

Biochemical signaling events, such as phosphorylation, acetylation, ubiquitylation, proteolytic cleavage, and so on are known to have mechanisms of activating or inactivating signaling proteins. Errors in signal transduction can lead to altered development and incorrect behavioral decisions which could result in abnormal end points of development. The relationships of signaling proteins are thought to be critical in determining cell behavior and maintaining cellular homeostasis, therefore, mis-regulation in the expression of genes and their regulators will be reflected on these cellular signaling networks which in turn lead to abnormal end points of development such as cancer and other diseases. miRNAs are posttranscriptional regulators, it is reasonable to think that miRNAs have great potential to regulate signaling networks.

Signaling networks are presented as graphs containing both directed and undirected links. In the networks nodes represent proteins, directed links represent activation or inactivation relationships between proteins, while undirected links represent simply physical interactions between proteins. Comparing to other types of cellular networks, signaling networks are far more complex in terms of the relationships between proteins (Wang et al. 2007).

Network-structural analysis of cellular signaling networks has been limited by the lack of comprehensive datasets for signaling networks. In the past few decades, enormous efforts have been made to study signaling pathways and generated lots of signaling information, especially in mammalian genomes. However, this historically generated information is scatted in literature. Recently, different researchers began to manually curate signaling information and organized them as signaling pathways such as EGFR signaling network and BioCarta signaling pathway database (http://www.biocarta.com/) (Oda and Kitano, 2006; Oda et al. 2005) or signaling networks (Awan et al. 2007; Ma'ayan et al. 2005). Other researchers used high-throughput technologies to fish new signaling proteins and their interactions based on large-scale experimental studies of protein interactions or genetic interactions between known signaling proteins and other proteins in genome (Barrios-Rodiles et al. 2005; Lehner et al. 2006). All these efforts offer new possibilities to analyze large and complex cellular signaling networks using the graph and network theory mentioned above.

So far only a few studies have been conducted for large-scale structural analysis of cellular signaling networks. In 2005, the first network-structural analysis of a literature-mined human cellular signaling network containing ~500 proteins was conducted, and showed that signaling pathways are intertwined in order to manage the numerous cell behavior outputs (Ma'ayan et al. 2005). This work provides a framework for our understanding of how signaling information is processed in cells. In 2006, we conducted an analysis of miRNA regulation of the human signaling network using the same dataset (Cui et al. 2006). In 2007, we collected more signaling information and extended the signaling proteins and their relations to the human signaling network. As a result, the new human signaling network contains more than 1,100 signaling proteins (Awan et al. 2007). Subsequent analysis of cancer-associated genes and cell mobility genes on the signaling network reveals the patterns of oncogenic regulation during tumorigenesis and finding oncogenic hotspots on the human signaling network (Awan et al. 2007). Notably, different regulatory patterns of oncogenic and cell mobility genes on the human signaling network have been observed (Awan et al. 2007). More recently, we further extended the human signaling network to contain more than 1,600 signaling proteins. Our efforts for curating signaling information include recording gene/protein names, molecular types (e.g., growth factor, ligand, adaptor, scaffold and so on), biochemical reactions (e.g., phosphorylation, acetylation, ubiquitylation and so on), and interaction types (e.g., activation, inhibition and simply physical interactions of proteins), cellular locations of the proteins and so on. These data are freely available on our website: http://www.bri.nrc.ca/wang/. We further conducted an integrative analysis of cancer casually implicated genetic and epigenetic alterations onto the human signaling network compiled from this more comprehensive dataset. Our analysis revealed where the oncogenic stimuli are embedded in the network architecture and illustrated the principles of triggering oncogenic signaling events by genetic and epigenetic alterations, furthermore, we extracted a human cancer signaling map and showed that different parts/regions of the cancer signaling map are required to coopt during tumorigenesis (Cui et al. 2007c). Using the same network, we performed a comprehensive analysis of the siganling network in an evoltionary context and underscored new insights into the evolution of cellular signaling networks (Cui et al. manuscript submitted). Collectively, these efforts indicate that integrative analysis of signaling networks with other datasets would highlight new insights into signaling mechanisms in different biological aspects, e.g., principles of miRNA regulation of signaling networks, cancer development/progression and evolution.

### 1.3.1.2 Strategies of miRNA regulation of cellular signaling networks

As miRNAs are able to directly and specifically knock down protein expression, we hypothesized that miRNAs might play an important role in the regulation of the strength and specificity of cellular signaling networks through directly controlling the concentration of network components (proteins) at post-transcriptional and translational levels. We took genome-wide computationally predicted miRNA target genes from two recent studies (Krek et al. 2005; Lewis et al. 2005) and then mapped all the overlapped miRNA targets onto the human signaling network proteins to conduct an network-structural analysis. This analysis revealed several strategies of miRNA regulation of signaling networks (Cui et al. 2006).

We found that miRNAs more frequently target signaling proteins than others, e.g., 29.4% vs 17% of the network proteins and the total genes in human genome are miRNA targets. This discovery implies that miRNAs might play a relatively more important role in regulating signaling networks than in other cellular processes. Normally, in signaling networks, cellular signal information flow initiates from extra-cellular space, a ligand binds to a cellular membrane receptor to start the signal, which is then transmitted by intracellular signaling components in cytosol and finally reaches the signaling components in the nucleus. We found that the fraction of miRNA targets increases with the signal information flow from the upstream to the downstream, e.g., from ligands, cell surface receptors, intracellular signaling proteins to nuclear proteins (Figure 1).

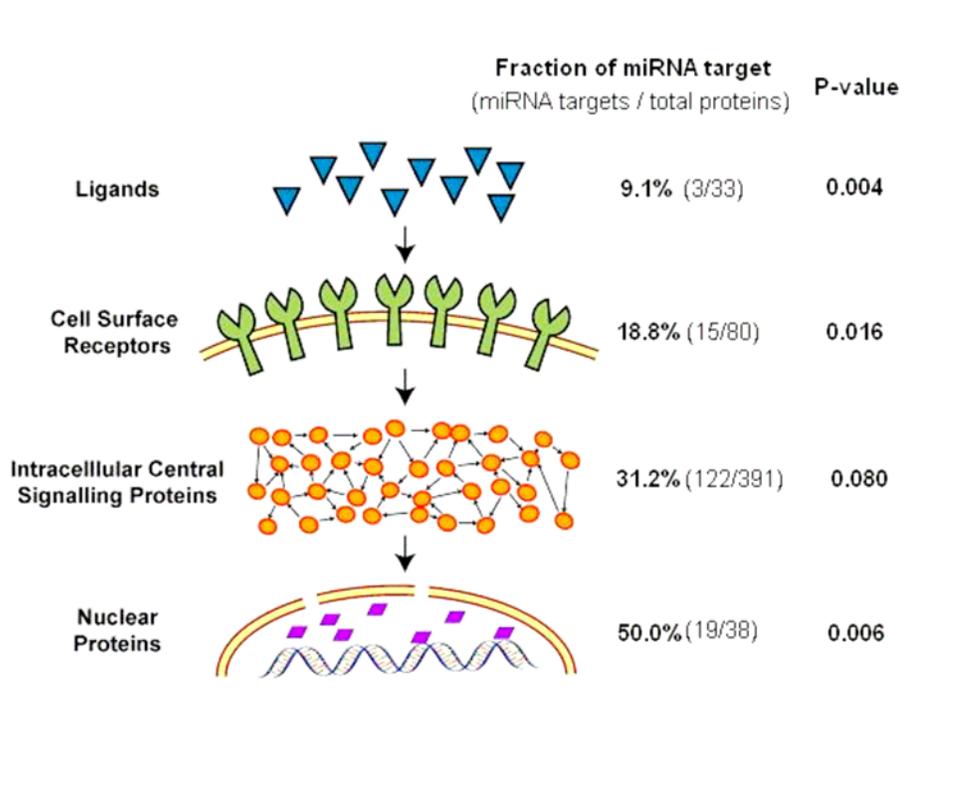

**Figure 1. Distribution of miRNA targets in a human signaling network at different signaling stages**. Signaling proteins are divided into four groups, e.g. ligands, cell surface receptors, intracellular central signaling proteins and nuclear proteins, according to their cellular locations in the signaling pathways. miRNA targets were mapped onto each group. miRNA target rate in each group was then calculated.

For example, only 9.1% of the ligands are miRNA targets, whereas half of the nuclear proteins, most of which are transcription factors, are miRNA targets. In other words, the miRNA targets are enriched more than five times in the most downstream proteins compared to the most upstream proteins. In signaling networks, adaptor proteins recruit downstream signaling components to the vicinity of receptors. They activate, inhibit or relocalize downstream components through direct protein-protein interactions. Adaptors do not have enzyme activity, but physically interact with upstream and downstream signaling proteins. One adaptor is able to recruit distinct downstream components in different cellular conditions. We found that miRNAs preferentially target the downstream components of adaptors, which have potential to recruit more downstream components. For example, the adaptor Grb2 directly interacts with 14 downstream signaling proteins, half of which are miRNA targets. These downstream components are functionally involved in different signaling pathways that lead to different cellular outputs. For example, SHC regulates cell growth and apoptosis through activation of small GTPases of the Ras family, while NWASP is involved in the regulation of actin-based cytoskeleton through activation of small GTPases of the Rho family. These two components are targeted by different miRNAs. To accurately respond to extracellular stimuli, adaptors need to selectively recruit downstream components. If an adaptor can recruit more downstream components, these components should have a higher dynamic gene expression behavior. This principle is in agreement with the fact that miRNAs have a high spatio-temporal expression behavior, suggesting that miRNAs might play an important role for precise selection of cellular responses to stimuli by controlling the concentration of adaptors' downstream components.

We further showed that miRNAs more frequently target positively linked network motifs and less frequently target negatively linked network motifs. A complex signaling network can be broken down into distinct regulatory patterns, or network motifs, typically comprised of three to four interacting components capable of signal processing (Babu et al. 2004; Barabasi and Oltvai, 2004). In our previous work, we showed that mRNA decay plays an important role in motif regulatory behavior (Wang and Purisima, 2005). In the network, we identified 11 types of motifs in the network (Figure 2). We classified each type of motif into several subgroups based on the number of nodes that are miRNA targets. For example, the three node network may have none of their nodes as a miRNA target (category 0), or may have just one of their nodes as a miRNA target (category 1), or 2 (category 2) or all three as miRNA targets (category 3). For each motif, we calculated the ratio of positive links to the total directional (positive and negative) links (termed as Ra) in each subgroup and compared it with the average Ra in all the motifs, which is shown as a horizontal line in Figure 3. For most motifs, the Ra in the subgroup in which none of the nodes are miRNA targets is less than the average Ra of all the motifs (Figure 3, $P < 4 \times 10^{-3}$, Wilcoxon Ranksum test). This result suggests that miRNAs less frequently target negative regulatory motifs. In contrast, for most motifs, the preponderance of positive links in the subgroups increased as the number of miRNA-targeted components rose (Figure 3). More significantly, when all nodes are miRNA targets in a motif, all the links in the motif are positive links ($P < 0.01$, Wilcoxon Ranksum test). These results suggest that miRNAs have high potential to target positively linked motifs. For example, AP1 (activator protein 1), CREB (cAMP-responsive element-binding protein) and CBP (CREB-binding protein) form a three-node positive feedback loop. All of the three proteins are miRNA targets.

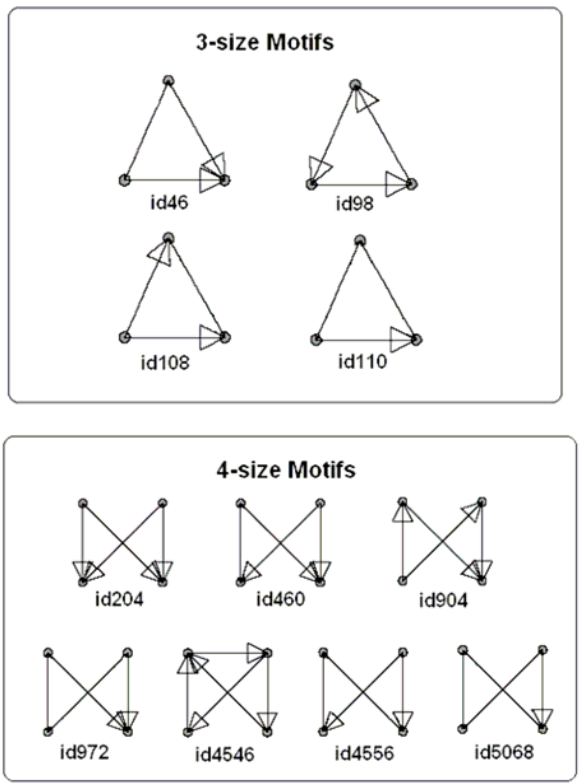

**Figure 2. Schematic diagram of the motifs identified within a human signaling network.**
Network motifs were identified using Mfinder program. The number under each motif is the motif ID number of the corresponding motif. Non-directional links represent neutral links and directional links represent positive links or negative links. The network motif ID numbering system is from Alon's motif dictionary (http://www.weizmann.ac.il/mcb/UriAlon/NetworkMotifsSW/mfinder/motifDictionary.pdf).

Positive feedback loops are often used to covert a transient signal into a long-lasting cellular response and make developmental switches (Cui et al. 2006; Ferrell, 2002). In the positive feedback loops, noise or fluctuation in any component can be easily amplified, and then driving the system to switch states randomly. In this situation, a negative control would enhance filtering or buffering such noise or fluctuation amplification. Randomly switching phenotypes has been observed in the yeast galactose network, which contains both positive and negative feedbacks. When the negative feedback was removed from the network, the genes in the network would randomly switch on and off over time (Acar et al. 2005). Compared to transcriptional repressors, miRNAs are likely to tune target protein levels more rapidly at posttranscriptional level. Thus, miRNAs could significantly shorten the response delay. Therefore, by regulating positive regulatory loops, miRNAs might provide fast feedback responses and more effective noise filtering as well as precise definition and maintenance of steady states. In another study, we showed that miRNA indeed buffers gene expression noise cross species (Cui et al. 2007b). Consider the facts that positive feedback circuits are abundant in genomes (Brandman et al. 2005; Ferrell, 2002), we surmise that miRNAs regulation of positive regulatory loops might provide a common mechanism for filtering noise and buffering. Compared to transcriptional repressors, miRNAs are likely to tune target protein levels more rapidly at posttranscriptional level.

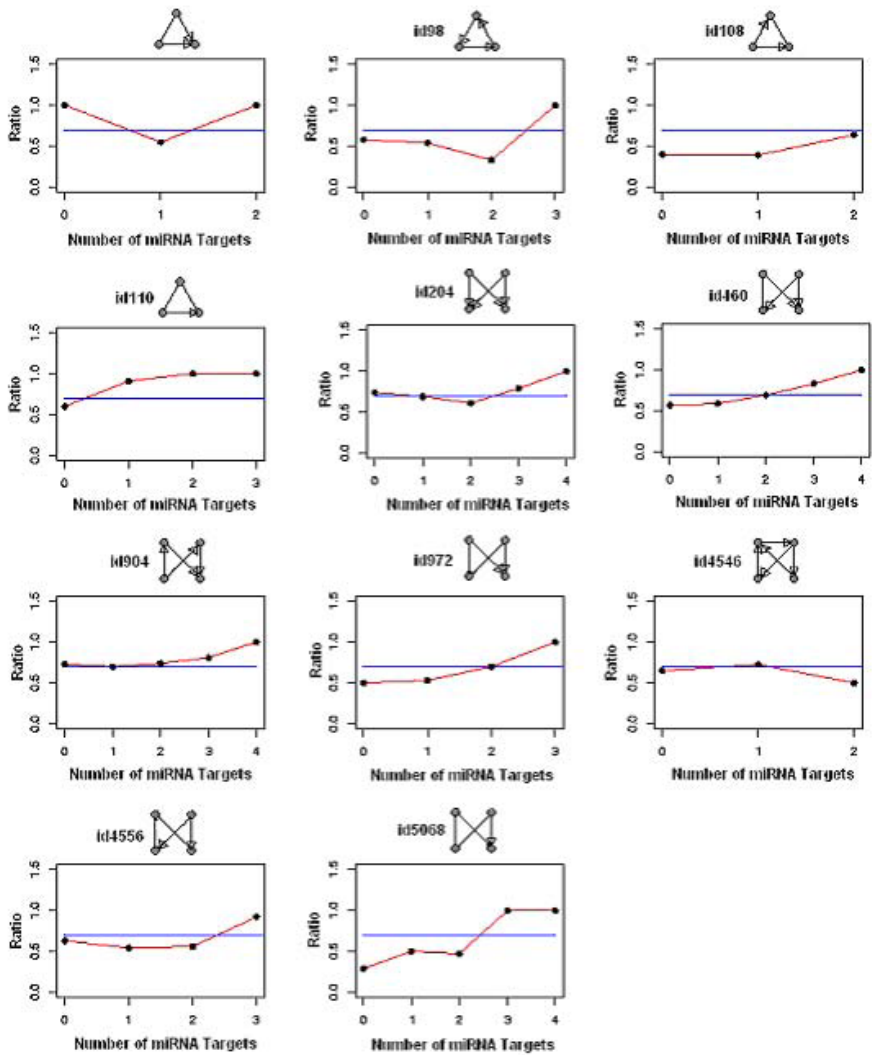

**Figure 3. Relative abundance of positive links in the individual subgroups of each type of network motif**.
Each type of motif was classified into several subgroups according to the number of nodes which are miRNA targets. For example, a three-node motif can be divided into four subgroups in which the miRNA target numbers are 0, 1, 2 and 3 respectively. The ratio of positive links to total positive and negative links in each subgroup was calculated and plotted as function of miRNA target numbers per motif. The horizontal lines indicate the ratio of positive links to the total positive and negative links in all of the respective network motifs. The network motif ID numbering system is from Alon's motif dictionary (http://www.weizmann.ac.il/mcb/UriAlon/NetworkMotifsSW/mfinder/motifDictionary.pdf).

To explore which cellular machines of the signaling network are regulated by miRNAs, we explored the relations between network themes and cellular machines. Network motifs are often linked together to form larger subgraphs. Network themes are examples of such larger subgraphs which are enriched topological patterns containing clusters of overlapping motifs, represent a higher order of regulatory relationships between signaling proteins and tie to particular biological functions (Zhang et al. 2005). To find the network themes which are regulated by each miRNA, we used the network motifs that contain at least one miRNA target and examined whether some of these motifs could aggregate into clusters. We found that in general, the network motifs regulated by each miRNA formed 1 or 2 network themes. The sizes of the network themes range from 4 to 145 nodes. Most of network themes contain more than 20 nodes. Statistical analysis of the associations between these network themes and cellular machines (transcription machinery, translation machinery, secretion apparatus, motility machinery and electrical response) revealed that nearly 60% of miRNAs in this study could be associated to one or more cellular machines of the signaling network.

We also uncovered that highly linked scaffold proteins have higher probability to be targeted by miRNAs. For example, CRK and SNAP25 are targeted by six miRNAs (miR-1, miR-10a, miR-126, miR-133a, miR-20 and miR-93) and five miRNAs (miR-1, miR-128a, miR-130a, miR-153 and miR-27b), respectively. Scaffold protein neutrally linked to other two proteins that are either positively or negatively connected (Figure 2, Network Motif id110). Unlike adaptors, scaffold proteins do not directly activate or inhibit other proteins but provide regional organization for activation or inhibition between other proteins. Scaffold proteins are able to recruit distinct sets of proteins to different pathways and thus maintain the specificity of signal information flows. Higher linked scaffold proteins can recruit more protein sets and have a higher degree of spatio-temporal expression behavior. The expression of miRNAs is highly specific for tissues and developmental stages, therefore it makes perfect sense that higher linked scaffold proteins are regulated by more miRNAs.

Finally we discovered that miRNAs avoid targeting common components of cellular machines in the network. In the network, we identified 70 proteins that are shared by all of the five basic cellular machines, e.g., transcription machinery, translation machinery, secretion apparatus, motility machinery and electrical response. We found that only 14.3% of the 70 proteins are miRNA targets, a significant under-representation compared to the fraction of miRNA targets (29.4%) in the network ($P < 2 \times 10^{-4}$). This result suggests that miRNAs avoid disturbing basic cellular processes, because these common proteins are highly shared by basic cellular machines and should be frequently used in various cellular conditions.

These rules or principles indicate that miRNAs regulate signaling networks in multiple ways. By selectively regulating positive regulatory motifs, highly connected scaffolds and the most network downstream components, miRNAs may provide a mechanism to terminate the preexisting messages and facilitate quick and robust transitions for responses to new signals. These functions fit the spatio-temporal behavior of miRNA expression. On the other hand, miRNAs less frequently target negative regulatory motifs, common proteins of basic cellular machines and upstream network components such as ligands. Although the accuracy of the miRNA target prediction methods has been well demonstrated by experimental validation of randomly selected targets, 12% of them could not be proved as real targets. Therefore, we performed the sensitivity analysis to test the potential effects of the errors on the robustness of the rules we discovered. We mimicked false positives by randomly adding extra 10% and 20% of network proteins, which are not predicted miRNA targets, to the target list, performed the same analysis and recalculated the P values. In addition, we also removed 10% and 20% of miRNA targets to determine the effect of false negatives. The results indicate that the trend remains unchanged by the addition of the false positives or false negatives. Therefore, the principles we obtained in this analysis are robust against substantial errors.

**1.3.2 miRNA regulation of gene regulatory networks**
Gene regulatory networks describe the regulatory relationships between transcription factors and/or regulatory RNAs and genes. Theoretically in a cell the entire gene regulatory network encodes a blueprint of gene regulatory relations and a framework for combinatorial mechanisms of using different regulatory relations to perform distinct biological functions. The network reflects the evolutionary selection, e.g., mRNAs of the hub transcription factors decay faster than other transcription factors (Batada et al. 2006; Wang and Purisima, 2005). In the past 50 years, *E. coli* and yeast have been used as model organisms to study gene regulation. Therefore, rich information about gene regulatory relations for these organisms has been documented in literature. RegulonDB, a manually curated database for collecting gene regulatory relations for *E. coli*, is one of the efforts to gather gene regulatory relations in literature (Salgado et al. 2004). A genome-wide determination of gene regulatory relations using chromatin immunoprecipitation coupled with DNA microarray in yeast represents an effort to uncover gene regulatory relations via high-throughput approaches (Lee et al. 2002). These efforts made it possible to analyze gene regulatory networks in a large-scale manner. Extensive analyses of gene regulatory networks in *E. coli* and yeast have been conducted, ranging from static network-structural analysis (Babu et al. 2004; Lee et al. 2002), network dynamic analysis (Babu et al. 2004; Luscombe et al. 2004), evolutionary analysis (Conant and Wagner, 2003), network decomposition (Shen-Orr et al. 2002) to integrative analysis (Wang and Purisima, 2005). However, gene regulatory information is still less comprehensive in mammalian cells.

To get insights into how miRNA interacts with gene regulatory networks in humans, we took a dataset which represents three transcription factors, OCT4, NANOG and SOX2 and their target genes in human embryonic stem cells (Boyer et al. 2005). The regulatory relationships between the transcription factors and their target genes were determined by using chromatin immunoprecipitation coupled with DNA microarray. The three transcription factors totally regulate 2,043 genes, of which 1,314 genes are co-regulated by two of the three TFs and 391 genes are co-regulated by all of the transcription factors. Using this dataset, we built a small gene regulatory network in which nodes represent transcription factors or genes, and links represent regulatory relations between transcription factors and the regulated genes or transcription factors (Cui et al. 2007a). We mapped the miRNA targets onto the genes of the network. For genes in the network, we divided them into three groups, in which they are regulated by one, two and all three of the transcription factors, respectively, and counted the number of genes that are miRNA targets and the number of genes that are not miRNA targets, respectively, in each group. We revealed that miRNA targets are significantly enriched in the genes that are regulated by more transcription factors. These results tell us that a gene that is regulated by a larger number of transcription factors is also more likely to be regulated by miRNAs.

Because the network is very small, e.g., containing only three transcription factors, we were not confident whether the conclusion we obtained above is robust. To validate and expand above observation, we examined the relationship between transcription factors and miRNAs for gene regulation at a genome-wide scale (Cui et al. 2007a). Although we were not able to get the datasets for a genome-wide gene regulatory network in human, we could access the datasets, which are computationally determined, of the number of transcription-factor–binding-site (TFBS) on the promoter region for each gene in human genome. TFBSs or *cis*-regulatory elements are normally located in the promoter region of a gene. Transcription factors regulate a gene through binding to the TFBSs of the gene. The two TFBS datasets were taken from recent publication of Cora et al. (Cora et al. 2005) and Xie et al. (Xie et al. 2005). Generally speaking, the more TFBSs a gene has, the more transcription factors the gene is regulated by, and the more complex its regulation can be as provided by various possible combinations of transcription factors.

We performed analysis to determine the relationship between the number of the TFBSs and the possibility to be a miRNA target of genes. Toward this end, we grouped genes based on their TFBS numbers (TFBS-count). The TFBS-count is significantly correlated with the miRNA target rate (Pearson's correlation coefficient $r = 0.94$, $P < 3.5 \times 10^{-68}$). For example, the miRNA target rate is doubled from the group of genes that have less than 10 TFBSs to those that have more than 100 TFBSs (from ~35% to ~70%). A similar result was obtained using the TFBS dataset from Xie et al. ($r = 0.97$, $P < 3.9 \times 10^{-113}$). These results are in agreement with the finding in the human stem cell gene regulation and therefore strongly suggest that miRNAs preferentially target the genes that bear more TFBSs has broad applicability. On the other hand, we analyzed the relationship between the number of miRNAs and the number of TFBSs in the same genes. We found a significant correlation ($r = 0.74$, $P < 6.1 \times 10^{-12}$). A similar result was obtained when using the TFBS dataset of Xie et al. ($r = 0.72$, $P < 9.5 \times 10^{-12}$). These results suggest that the genes that are targeted by more miRNAs have more TFBSs (Cui et al. 2007a).

Collectively, we uncovered a basic rule of miRNA regulation of gene regulatory networks: a gene that is regulated by more transcription factors is also more likely to be regulated by miRNAs. These results indicate that the complexity of gene regulation by miRNAs at the post-transcriptional level is positively related to the complexity of gene regulation by transcription factors at the transcriptional level in human genome.
Genes, which are more complexly regulated at transcriptional level, are required to be turned on more frequently, furthermore, are more likely to be expressed at different temporal and spatial conditions, therefore, they are also required to be turned off more frequently. miRNAs as negative regulators can exert the turning-off function at post-transcriptional level through repressing mRNA translation and/or mediating cleavage of mRNAs. This is a potentially novel discovery of mechanism for coordinated regulation of gene expression. Such coordinately regulated genes are enriched in certain biologocal processes and functions, particularly in those involved in developmental processes.

In a seperate study, we showede that miRNAs preferientially regulate positive regulatory loops of signaling networks (Cui et al. 2006). It is also reasonable to hypothesize that miRNAs preferientially regulate positive regulatory loops of gene regulatory networks. Given that positive feedback circuits are abundant in genomes (Brandman et al. 2005; Ferrell, 2002), we surmise that miRNAs frequently regulate gene regulatory networks by targeting positive regulatory loops. Unfortunately the datasets for positive and negative regulatory relations are currently unavailable in humans/rodents and even in worm and fly, and so we are not able to test this hypothesis at this moment.

### 1.3.3 miRNA regulation of metabolic networks

Metabolites are critical in a cell. Certain metabolites are the basic building blocks of proteins, DNAs and RNAs, some metabolites such as fatty acids take part in the cellular processes for growth, development and reproduction, while some others are involved in defense mechanisms against parasites and cell signaling. Biochemical characterization of metabolic reactions and enzymes has been conducted for many years. Traditionally, metabolic reactions are organized and illustrated as metabolic pathways. In term of pathway components, metabolic pathways are so far the clearest and the most comprehensive. Genome sequencing efforts offer comparative genomic analysis of metabolic reactions and enzymes cross many species. As a result, the information for metabolic pathways is more enriched than before. It makes it possible to build comprehensive metabolic maps at this time (Feist et al. 2007).

Many metabolites are shared by different metabolic pathways and are further intertwined to form a complex metabolic network. Thus, various cellular activities are accompanied with the changes of metabolism. It is essential to control the rates of metabolic processes in response to changes in the internal or external environment for living cells. Mechanisms that control metabolic networks are complex and involve transcriptional, post-transcriptional and translational regulations. For a long time, we have reasons to believe that the enzymes of metabolic networks are tightly controlled by transcription factors. Moreover, the principles of transcriptional regulation of metabolic networks by transcription factors have been illustrated through an integrative analysis of gene expression profiles and the yeast metabolic network (Ihmels et al. 2004). Since miRNAs have emerged as an abundant class of negative regulators, it is reasonable to think that miRNAs might extensively regulate metabolic networks. Indeed, miRNAs have been shown to regulate amino acid catabolism, cholesterol biosynthesis, triglyceride metabolism, insulin secretion, and carbohydrate and lipid metabolism, although the molecular mechanisms of miRNA regulation of metabolism are not clear (Krutzfeldt and Stoffel, 2006).

We systematically analyzed the human and *D. melanogaster* metabolic networks by integrating miRNA target genes onto the networks (Tibiche et al. manuscript submitted).

In both networks, miRNAs selectively regulate certain metabolic processes such as amino acid biosynthesis, certain sugar and lipid metabolisms, so that they can selectively control certain metabolite production. When miRNAs regulate specific individual metabolic pathway, they often regulate the last reaction step (LRS) of that pathway. Furthermore, once miRNAs regulate the LRS of a pathway, the cut vertex to the LRS and other enzymes that are in the upstream metabolic flows to the LRS are also enriched with miRNA targets. A cut vertex or a cutpoint is such a bottleneck node that its deletion will disconnect at least one component from the network. Cut vertexes are in crucial network positions and become bottlenecks of the network, and therefore control metabolic flows from a part to another in the network. These results imply that miRNA is strongly involved in coordinated regulation of metabolic processes in metabolic networks.

**1.3.4 miRNA regulation of protein interaction networks**
Protein interaction networks provide a valuable framework for a better understanding of the functional organization of the proteome and offer a mechanistic basis for most biological processes in organisms. Large-scale determination of interactions between proteins have been conducted in yeast, *E. coli* and other bacteria, worm, fly and humans (Butland et al. 2005; Gavin et al. 2002; Brechot et al. 1980; Rain et al. 2001; Stelzl et al. 2005; Rual et al. 2005; Giot et al. 2003; Ito et al. 2001; Li et al. 2004). Because the datasets for protein interaction networks are relatively easier to be accessed, extensive analyses of protein interaction networks have been conducted ranging from pure network structural analysis to the analyses of network motifs, network themes, network communities and evolution.

Recently, Liang and Li investigated the miRNA regulation of protein interaction networks (Liang and Li, 2007). miRNAs preferentially regulate the proteins which have more interacting partners in the network. Protein connectivity in a human protein interaction network is positively correlated with the number of miRNA target-site types. In principle, if a protein has more interacting protein partners, it normally takes part in more biological processes and then its expression is more dynamic (Wang et al. 2007). Therefore, it makes sense that when a protein has more interacting protein partners, it will be regulated by more transcription factors and more miRNAs. This is in agreement with our previous findings that miRNAs preferentially regulate the genes which are regulated by more transcription factors in the gene regulatory network (Cui et al. 2007a). Consistently, genes of two interacting proteins tend to be under similar miRNA regulation, which again is in agreement with these facts that genes encoding interacting proteins tend to have similar mRNA expression profiles (Li et al. 2004; Rual et al. 2005; Wuchty et al. 2006). Furthermore, our analysis of microarray profiles of miRNA target genes in different tissues showed a similar trend: broadly expressed mRNAs tend to be regulated by more miRNAs (Yu et al. 2007a). Highly linked proteins, e.g., hub proteins, can be divided into two groups based on clustering coefficient, which is defined as the fraction of the real number of links among a node's neighbors and the maximum possible number of links among them. A hub protein having a high clustering coefficient is likely to be an intra-modular hub, which interacts with most of its partners simultaneously to form a protein complex and completes a coherent function. On the other hand, a hub protein having a low clustering coefficient tends to be an inter-modular hub, which tends to interact with other proteins in different time and place, and then coordinates different functional modules (Liang and Li, 2007). It is understandable that inter-modular hub proteins are more likely to be regulated by miRNAs.

Notably, analyses of miRNA regulation of gene regulatory networks and the protein interaction networks obtained a consistent conclusion, which reflects a general rule of miRNA regulation of whole genome genes, because general networks (gene regulatory networks and the protein interaction networks) collect information for all kinds of activities in cell. On the other hand, analyses of miRNA regulation of metabolic networks and signaling networks obtained different results. This observation might reflect that cellular specific networks (metabolic networks and signaling networks) have local and specific cellular themes, while general networks (gene regulatory networks and the protein interaction networks) encode global features of cells.

### 1.3.5 miRNA regulatory network motifs

At present, experimentally large-scale studies of miRNA regulation of various cellular networks are still not trivial. However, it is experimentally feasible for characterizing miRNA regulatory network motifs, or miRNA regulatory circuits. As we mentioned above, network motifs are the statistically significant recurring structural patterns or small subgraphs or sub-networks, which are a result of convergent evolution at the network level, and carry out important functions in cells. Several studies have experimentally explored miRNA regulatory motifs. For example, the secondary vulva cell fate in *C. elegans* is promoted by Notch signaling, which also activates miR-61, which in turn posttranscriptionally represses an inhibitory factor of Notch signaling, thereby stabilizing the secondary vulva fate (Yoo and Greenwald, 2005). Similar circuits are also found in the differentiation of neurons in *C. elegans* (Johnston, Jr. et al. 2005), eye development in *Drosophila* (Li and Carthew, 2005; Li et al. 2006), and granulocytic differentiation in human (Fazi et al. 2005). Another example of miRNA regulatory circuits is that miR-17-5p represses E2F1, and both are transcriptionally activated by c-Myc in human cells (O'Donnell et al. 2005). More recently, two groups of researchers published data showing that miR-34 family is a key component of the p53 tumor suppressor network, which controls cellular responses to signals such as DNA damage and oncogene activation (He et al. 2007; Raver-Shapira et al. 2007).

To systematically characterize miRNA regulatory circuits in human and mouse genomes, one computational study tended to take the advantages of intron-based miRNAs, or embedded miRNAs (Tsang et al. 2007). More than 80% of all known miRNAs in human and mouse are embedded in introns of coding or noncoding genes (Kim and Kim, 2007; Rodriguez et al. 2004). One could hypothesize that the intron-based miRNAs might be co-expressed with their host genes. Indeed, the expression profiles of most embedded miRNAs examined thus far are highly correlated to their host genes at both the tissue and individual cell levels (Aboobaker et al. 2005; Baskerville and Bartel, 2005; Li and Carthew, 2005), suggesting that they tend to be cotranscribed at identical rates from the same promoters (Kim and Kim, 2007). Theses facts led to the assumption that the relative level of host-gene transcription across conditions can accurately serve as a proxy for that of the embedded miRNAs.

An upstream regulatory factor triggers the expression of an embedded miRNA and its targets at the same time, while the transcribed miRNA also regulate its targets. Intuitively, two types of miRNA regulatory motifs, or feedforward loops can be built up (Figure 4a, b): Type II miRNA feedforward loops, in which an upstream factor could repress the transcription of a target gene and simultaneously activate the transcription of a miRNA that inhibits target gene translation, and Type I miRNA feedforward loops, in which an upstream factor activates the transcription of a target gene and simultaneously activates the transcription of a miRNA that inhibits target gene translation. One example has been characterized experimentally for Type I miRNA feedforward loops, where miR-17-5p represses E2F1, and both are transcriptionally activated by c-Myc in human cells (O'Donnell et al. 2005). This kind of loop has the potential to provide a host of regulatory and signal processing functions (Hornstein and Shomron, 2006).

By analyzing the embedded miRNAs in human and mouse using the Novartis human and mouse expression atlas comprising 61 tissues/cell types (Su et al. 2004), Type II miRNA feedforward loops have been found to be prevalent for a significant fraction of the embedded miRNAs. This result is in agreement of previous findings that predicted target transcripts of several tissue-specific miRNAs tend to be expressed at a lower level in tissues where the miRNAs are expressed (Farh et al. 2005; Stark et al. 2005; Yu et al. 2007a). Furthermore, a significant fraction of Type I miRNA feedforward loops were also found, especially prevalent in mature neurons.

In Type II miRNA feedforward loops, a miRNA regulates its targets coherently with transcriptional control, thereby reinforcing transcriptional logic at the posttranscriptional level. As suggested, such circuits can serve as a surveillance mechanism to suppress "leaky" transcription of target genes (Hornstein and Shomron, 2006; Stark et al. 2005). It is reasonable to think that these loops would act in concert with other regulators to increase the feedback strength and enhance the robustness of irreversible cellular differentiation. On the other hand, Type I miRNA feedforward loops might prevent noise-driven transitions into proliferation as illustrated in the example of c-Myc/E2F1/miR-17-92 network (Sylvestre et al. 2007; O'Donnell et al. 2005).

Another effort to computationally identify miRNA network motifs has been conducted by analyzing the cooperation between transcription factors and miRNAs for regulating the same target genes (Shalgi et al. 2007). The computational techniques for this kind of analysis are similar to those used in identifying transcription factor binding sites. They first looked for transcription factor-miRNA pairs with a high rate of co-occurrence in the promoters and 3'UTR of the regulated genes. Statistical tests showed that transcription factor-miRNA pairs significantly co-occur, which is in agreement with the findings of miRNA regulation of gene regulatory networks. Secondly, they searched miRNA networks using randomization tests. Similar to above report, Type I and II feedforward loops are also discovered. In addition, two other types of miRNA network motifs were also documented: composite regulatory loops and indirect feedforward loops (Figure 4c, d). In the former motif, a miRNA (MR) represses a transcription factor (TF) and a target gene (G), while the transcription factor (TF) activates the miRNA (M) and the target gene (G). In the latter motif, a transcription factor, TF1 activates another transcription factor, TF2 and a target gene, G, in turn, the TF2 activates a miRNA, M, which represses the target gene, G. However, the functions of these two motif types are not clear yet.

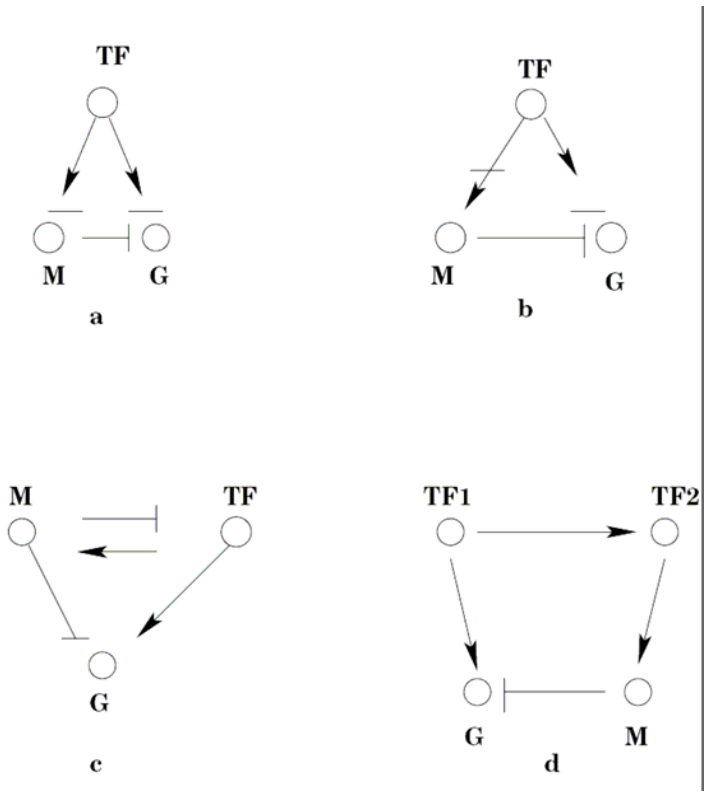

**Figure 4. Types of miRNA regulatory network motifs in the human genome.**
M represents miRNA while TF and G represent transcription factors and targeted genes, respectively. Links represent regulatory relations. Arrows represent activation, while a bar represents inhibition. **(a)** Type I miRNA feedforward motif, in which TF sends either activation signal or inhibition signal at the same time to M and G. **(b)** Type II miRNA feedforward motif. In one setting, TF activates M and inhibits G at the same time, while in another setting, TF inhibits M and activates G. **(c)** Composite regulatory motif. **(d)** Indirect feedforward motif.

**1.4 Summary**
In summary, miRNAs are extensively involved in gene regulation as network motifs in genomes. By analyzing the interactions between miRNAs and general networks (gene regulatory and protein interaction networks), a common miRNA regulatory principle is emerging: miRNAs preferentially regulated the genes that have high regulation complexity. This fact suggests a novel mechanism of coordinated regulation between transcriptional level and posttranscriptional level for gene regulation. In addition, for cellular specific networks (metabolic and signaling networks), miRNAs have different regulatory strategies. For example, miRNAs preferentially regulate positive regulatory motifs, highly connected scaffolds and the most network downstream components of cellular signaling networks, which might provide a mechanism to terminate the preexisting messages and facilitate quick and robust transitions for responses to new signals. On the other hand, miRNAs less frequently target negative regulatory motifs, common proteins of basic cellular machines and upstream network components such as ligands in signaling networks. In metabolic networks, miRNAs selectively regulate the genes which have specific network structural features on the network, which might provide effective and selective regulation of cellular metabolism.

**Acknowledgements**
This work is partially supported by Genome and Health Imitative. I thank our team members, Dr. Y. Deng and Mrs. M. Mistry for comments.